\documentclass[useAMS,usenatbib]{mn2e}
\usepackage{graphicx}

\title[Multi-frequency detection of point sources in CMB maps]{A novel
multi-frequency technique for the detection of point sources in Cosmic
Microwave Background maps} \author[Herranz et
al.]{D. Herranz$^{1}$\thanks{E-mail: herranz@ifca.unican.es},
M. L\'opez-Caniego$^{2}$, J.~L. Sanz$^{1,3}$ and
J. Gonz\'alez-Nuevo$^{4}$ \\ $^{1}$ Instituto de F\'\i sica de
Cantabria, CSIC-UC, Av. los Castros s/n, Santander, 39005, Spain \\
$^{2}$ Astrophysics Group, Cavendish Laboratory, J.J. Thomson Avenue,
CB9 0E1, Cambridge, United Kingdom \\ $^{3}$ CNR Istituto di Scienza e
Tecnologie dell'Informazione, via G. Moruzzi 1, I-56124, Pisa, Italy
\\ $^{4}$ SISSA-I.S.A.S., via Beirut 4, I-34014 Trieste, Italy}

\begin{document}

\date{Received --, Accepted --}

\pagerange{\pageref{firstpage}--\pageref{lastpage}} \pubyear{2008}

\maketitle

\label{firstpage}

\begin{abstract}

In this work we address the problem of simultaneus multi-frequency
detection of extragalactic point sources in maps of the Cosmic
Microwave Background. We apply a new linear filtering technique, the
so called `matched matrix filters', that incorporates full spatial
information, including the cross-correlation among channels, without
making any a priori assumption about the spectral behaviour of the
sources. A substantial reduction of the background is achieved thanks
to the optimal combination of filtered maps. We describe in detail the
new technique and we apply it to the detection/estimation of radio
sources in realistic all-sky \emph{Planck} simulations at 30, 44, 70
and 100 GHz. Then we compare the results with the mono-frequencial
approach based on the standard matched filter, in terms of reliablity,
completeness and flux accuracy of the resulting point source
catalogs. The new filters outperform the standard matched filters for
all these indexes at 30, 44 and 70 GHz, whereas at 100 GHz both kind
of filters have a similar performance. We find a noticeable increment
of the number of true detections for a fixed reliability level. In
particular, for a $95\%$ reliability we practically double the number
of detections at 30, 44 and 70 GHz.

\end{abstract}


\section{Introduction} \label{sec:intro}

From the search of extrasolar planets to the study of active galactic
nuclei, one of the most common task in all the branches of Astronomy
is the detection of faint pointlike objects. Such objects have angular
sizes that are smaller than the angular resolution of the telescopes
that are used to observe them, and therefore they are usually referred
to as \emph{point sources}. 

A case of particular interest is the detection of extragalactic point
sources (EPS) in maps of the Cosmic Microwave Background (CMB). EPS
are known to be a relevant source of contamination for CMB studies,
specially at small angular scales, where they hamper the estimation of
CMB angular power spectrum both in temperature
\citep{tof98,zotti99,hob99,zotti05} and in polarization
\citep{tucci04,tucci05}. Therefore, for the sake of CMB analysis it is
necessary to detect and remove as many extragalactic point souces as
possible.

Moreover, in the frequency range spanned by CMB experiments the
properties of EPS are poorly studied. Only very recently the
\emph{Wilkinson Microwave Anisotropy Probe} (WMAP)
satellite~\citep{wmap0} has permitted the obtention of the first
all-sky complete point source catalogs above $\sim0.8$--$1$ Jy in the
23--94 GHz range of frequencies
\citep{wmap0,hinshaw07short,NEWPS07,chen08,wright08short}. The
upcoming \emph{Planck} mission~\citep{planck_tauber05} will allow us
to extend these catalogs down to lower flux limits and up to 857
GHz. There is interesting physics to be probed in this frequency
range. The new EPS catalogs provided by next generation CMB
experiments will not only allow us to follow the behaviour of source
counts from existing catalogs to microwave frequencies, but also to
study source variability and to discover rare objects such as inverted
spectrum radio sources, extreme gigahertz peaked spectrum (GPS)
sources and high-redshift dusty galaxies (see for example the
\emph{Planck Bluebook}~\citep{bluebook} for a brief, yet comprehensive
review of the rich phenomenology of EPS at microwave
frequencies). Thus, the task of detecting point sources is important
not only from the point of view of CMB science but also from the point
of view of extragalactic Astronomy as well.

Let us consider a single image taken at a given wavelength. Then the
problem consists on how to detect a number of objects, all of them
with a common waveform that is generally considered to be well known
(basically, the shape of point sources is that of the beam) but with
unknown positions and intensities, that are embedded in additive noise
(not necessarily white). In the field of microwave Astronomy, wavelet
techniques~\citep{vielva01,vielva03,MHW206,wsphere,NEWPS07}, matched
filters~\citep[MF,][]{tegmark98,barr03,can06} and other related linear
filtering
techniques~\citep{sanz01,naselsky02,herr02b,herr02c,can04a,can05a,can05b}
have proved to be useful. The common feature of all these techniques
is that they rely on the prior knowledge that the sources have a
distinctive spatial behaviour (i.e.~a known spatial profile, plus the
fact that they appear as compact objects as opposed to `diffuse'
random fields) that helps to distinguish them from the
noise. Detection can be further improved by including prior
information about the sources, i.e. some knowledge about their
intensity distribution, in the frame of a Bayesian
formalism~\citep{hob03,psnakesI}.

Most of the current and planned CMB experiments are able to observe
the sky at several wavelengths simultaneously. Multi-wavelength
information makes it possible to separate different astrophysical
components (as for example CMB from Galactic synchrotron emission)
that have different spectral behaviour. Although multiwavelength
component separation techniques have been very succesful in separating
diffuse components~\citep[for a recent comparative review of several
  methods applied to sky simulations very similar to the ones we will
  use in this paper, see][]{challenge08},
the detection of EPS has been usually attempted on a channel by
channel basis. The reason for this is that EPS form a very
heterogeneous population, constituted by a large number of objects
with very different physical properties, and therefore it is
impossible to define a common spectral behaviour for all of them. The
whole situation remains somewhat unsatisfactory: on the one hand, the
channel-by-channel approach based on the spatial behaviour works fine,
but a valuable fraction of the information that multi-wavelenght
experiments can offer is wasted this way. On the other hand, standard
component separation techniques based on spectral diversity have
problems when dealing with the very heterogeneous EPS components.

Thus, multi-wavelenght detection of EPS in CMB images remains a
largely unexplored field. In recent years, some attempts have been
done in this direction. For example, \citet{naselsky02b} combined
simulated multi-wavelenght maps ir order to increase the average
signal to noise ratio of point sources. In a similar
way,~\citet{chen08} and~\citet{wright08short} use combinations of the
\emph{WMAP} W and V bands in order to produce a CMB-free map in which
to better detect the elusive radio galaxies. Note that, in any case,
combined 'clean' maps are suitable for detecting more sources but not
for performing accurate photometry, unless the spectral index of all
the sources is known in advance.

An intermediate approach is to design filters that are able to find
compact sources thanks to their distinctive spatial behaviour while at
the same time do incorporate some multiwavelength information, without
pretending to achieve a full component separation and without assuming
a specific spectral behaviour for the sources. Very recently, the
authors have proposed a new technique, based on the so-called
\emph{`matched matrix filters'} \citep[MTXF,][]{herranz08a}, that goes
in this direction. The basic underlying ideas of the new method are:
\begin{itemize}
\item When a source is found in one channel, it will be also present
  in the same position in all the other channels.
\item The spatial profile of the sources may differ from channel to
  channel, but it is a priori known.
\item The second order statistics of the background in which the
  sources are embedded is well known or it can be directly estimated
  from the data by assuming that point sources are sparse. This
  knowledge about the second order statistics (namely, the
  background's power spectrum for each channel and the its
  correlations among the different channels) will be used to increase
  the signal to noise ratio of the sources.
\item We want to perform accurate photometry of the sources at every
  one of the frequencies covered by the experiment, independently of
  what is the spectral behaviour of any source in particular. 
\end{itemize} 
In~\citet{herranz08a} the authors presented the new methodology and
demonstrated its potential utility with a few toy simulations. In this
paper, we will study its applicability to real CMB experiments by
applying it to realistic simulations of the whole sky as will be
observed by the upcoming \emph{Planck} mission. We will focus on the
particular case of the detection of radio sources in the four lower
frequency \emph{Planck} channels (33--100 GHz), comparing the
performance of the new filters with the performance of the
well-stablished standard matched filters. In section~\ref{sec:matrixf}
we will summarize the foundations of the theoretical formulation of
matched matrix filters. In section~\ref{sec:toplanck} we will describe
the \emph{Planck} simulations that we use to test the method and we
will outline the main features of the code we have developed for its
implementation. The results of the exercise will be commented in
section~\ref{sec:results}. Finally, in section~\ref{sec:conclusions}
we will draw some conclusions.

\section{Matched matrix filters} \label{sec:matrixf}

The derivation of the matched matrix filters is fully described
in~\citet{herranz08a}\footnote{For economy, in the following we will
occasionally refer to the matched matrix filters just as `matrix
filters'}. However, for reasons of clarity, we will reproduce here the
main ideas of that paper here.

\subsection{Data model} \label{sec:data_model}

Let us consider a set of $N$ two-dimensional images (channels) in
which there is an unknown number of point sources embedded in a mixture
of instrumental noise and other astrophysical components. Without loss
of generality, let us consider the case of a single point source
located at the origin of the coordinates. Our data model is
\begin{equation} \label{eq:model}
  D_k \left( \vec{x} \right) = s_k \left( \vec{x} \right) +
  n_k \left( \vec{x} \right),
\end{equation}
where the subscript $k=1,\ldots,N$ denotes the index of the
image. The term $s_k (\vec{x})$ denotes the point source,
\begin{equation} \label{eq:ps}
  s_k \left( \vec{x} \right) = A_k \tau_k \left( \vec{x} \right),
\end{equation}
\noindent
where $A_k$ is the unknown amplitude of the source in the
$k^{\mathrm{th}}$ channel and $\tau_k (\vec{x})$ is the spatial
profile of the source (which is assumed to be known) and satisfies the
condition $\tau_k ( \vec{0} ) = 1$. The term $n_k (\vec{x})$ in
equation (\ref{eq:model}) is the generalized noise in the
$k^{\mathrm{th}}$ channel, containing not only instrumental noise, but
also CMB and all the other astrophysical components apart from the
point sources.  Let us suppose the noise term can be characterized by
its cross-power spectrum:
\begin{equation} \label{eq:noise_ps}
\langle n_k \left( \vec{q} \right) n^{*}_l \left( \vec{q}^{~\prime}
\right) \rangle = P_{kl} \left( \vec{q} \right) \delta^2 \left( \vec{q} -
\vec{q}^{~\prime} \right),
\end{equation}
where $\mathbf{P} = (P_{kl})$ is the cross-power spectrum matrix and
the symbol $*$ denotes complex conjugation. From now on, we assume
that the noise has zero mean:
\begin{equation}
  \langle n_k \left( \vec{x} \right) \rangle = 0.
\end{equation}

\subsection{Filtering with matrices of filters} \label{sec:filtering}

Since we are interested in doing accurate photometry in each one of
the $N$ available channels, we are bound to produce $N$ different
processed maps. Therefore, we are looking for a transformation that
starts with $N$ input channels and ends with other $N$ processed maps
where a) point sources are easier to detect and b) the amplitudes
$A_k$ are preserved. Besides, since we intend to use some
multi-wavelength information, we are interested in making all the $N$
input channels intervene in the elaboration of the any one of the
output maps. One possibility is to define a set of $N\times N$ filters
$\Psi_{kl} (\vec{x})$ such that the $N$ combined quantities
\begin{eqnarray} \label{eq:filtered_field}
  w_k (\vec{x}) & = & \sum_l \int d\vec{x}^{~\prime} \Psi_{kl} \left(
  \vec{x} - \vec{x}^{~\prime} \right) D_l \left( \vec{x}^{~\prime}
  \right) \nonumber \\ & = & \sum_l \int d \vec{q} \, \, e^{-i \vec{q}
  \cdot \vec{x}} \, \Psi_{kl} \left( \vec{q} \right) D_l \left(
  \vec{q} \right)
\end{eqnarray}
\noindent
are our processed maps. The last term of the equation is just the
expression of the filterings in Fourier space, being $\vec{q}$ the
Fourier mode and $\Psi_{lk}(\vec{q})$ and $D_l(\vec{q})$ the Fourier
transforms of $\Psi_{kl}(\vec{x})$ and $D_l(\vec{x})$, respectively.

We intend to use the combined filtered image $w_k (\vec{x})$ as an
estimator of the source amplitudes $A_k$ for all $k=1,\ldots,N$. Thus,
the filters $\Psi_{kl}$ must satisfy the condition that the
$k^{\mathrm{th}}$ filtered, combined image at the position of the
source is, on average over many realizations, an \emph{unbiased}
estimator of the amplitude of the $k^{\mathrm{th}}$ source. The other
requirement we will ask the processed maps is that the signal to noise
ratio of the sources is increased with respect to the input maps. In
other words, we want estimators $w_k$ to be not only unbiased, but
\emph{efficient} as well. Therefore, we need to minimize the variance
$\sigma_{w_k}$ of the combined filtered image.

The set of filters that minimize the variance $\sigma_{w_k}$ for all
$k$ while keeping the individual amplitudes $A_k$ constant for all the
point sources, independently of their frequency dependence, can be
shown to be given by the matrix equation:
\begin{equation} \label{eq:matrix_filters_eqs}
\mathbf{\Psi}^*  =  \mathbf{F} \mathbf{P}^{-1},
\end{equation}
\noindent
where
\begin{equation}
  \mathbf{F}  =  (F_{kl}),\,\,\mathbf{P} = (P_{kl}),
  \mathbf{\lambda} = (\lambda_{kl}),\,\,\mathbf{H} = (H_{kl}),
\end{equation}
\noindent
and where
\begin{eqnarray} \label{eq:matrix_filters_eqs2}
  F_{kl} & = &\lambda_{kl} \tau_l, \nonumber\\
  \mathbf{\lambda} & = & \mathbf{H}^{-1}, \nonumber \\ 
  H_{kl} & = & \int d\vec{q} \, \, \tau_k \left(\vec{q}\right) P^{-1}_{kl}
  \tau_l \left(\vec{q}\right) .
\end{eqnarray}

The set of filters we have developed naturally assume a structure that
is best expressed in the form of a matrix equation, hence the
denomination of \emph{`matched matrix filters'}.

\subsection{Properties of matched matrix filters and some particular cases}

\subsubsection{A single image} \label{sec:one}

If the number of images is $N=1$ it is easy to show that the matrix of
filters contains a single element, which is the complex conjugate of
the standard matched filter. For circularly symmetric source profiles,
the filter is real-valued and the resulting filter is exactly the same
as the standard matched filter.

\begin{figure}
  \includegraphics[width=\columnwidth]{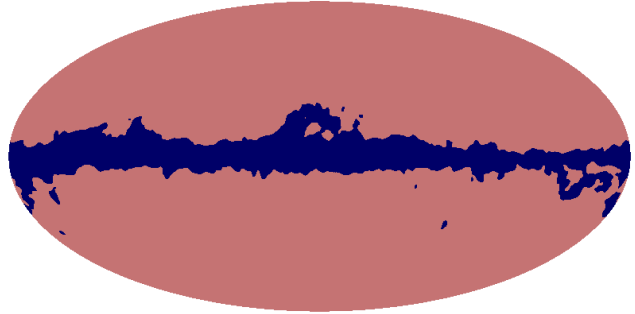}
  \caption{Mask used for the analysis. The mask covers
  $14.91\%$ of the sky, including the Galactic Plane plus some other
  densely contamined areas of the sky such as the Magellanic Clouds,
  the Ophiuchus Complex and the Orion/Eridanus Bubble.
  \label{fig:mask}}
\end{figure}

\subsubsection{Uncorrelated noise} \label{sec:uncorr}

From equations (\ref{eq:matrix_filters_eqs}) and
(\ref{eq:matrix_filters_eqs2}) it is straightforward to show that for
the particular case where the noise is totally uncorrelated among
channels the matrix of filters defaults to a diagonal matrix whose
non-zero elements are the complex conjugates of the standard matched
filters that correspond to each input channel. When the source
profiles are circularly symmetric
the filters are real-valued and the whole
process is equivalent to filter each channel independiently with the
appropiate matched filter.

\subsubsection{The $2\times 2$ case} \label{sec:2x2}

In this case, the explicit form of the MTXF is
\begin{eqnarray} \label{eq:2x2}
  \Psi^*_{11} & = & \frac{1}{N\Delta}[\phi_1 -
    \phi_2\frac{P_{12}b_{12}}{P_{11}b_{22}}], \nonumber \\ \Psi^*_{12} &
  = & \frac{1}{N\Delta}[-\phi_1\frac{P_{12}}{P_{22}} +
    \phi_2\frac{b_{12}}{b_{11}}], \nonumber \\ \Psi^*_{22} & = &
  \frac{1}{N\Delta}[\phi_2 - \phi_1\frac{P_{12}b_{12}}{P_{22}b_{11}}],
  \nonumber \\ \Psi^*_{21} & = &
  \frac{1}{N\Delta}[\phi_1\frac{b_{12}}{b_{22}} -
    \phi_2\frac{P_{12}}{P_{11}}], \nonumber \\ N & \equiv & 1 -
  \frac{b^2_{12}}{b_{11}b_{22}}, \nonumber \\ \Delta & \equiv & 1 -
  \frac{P^2_{12}}{P_{11}P_{22}}, \nonumber \\ b_{ij} & \equiv & \int
  d\,\vec{q}\frac{\tau_i\tau^*_j}{\Delta}\frac{P_{ij}}{P_{ii}P_{jj}},
\end{eqnarray}
\noindent where $i,j = 1, 2$ and $\phi_i$ are closely related to the
standard matched filters $\phi_i^{MF}$:
\begin{eqnarray} \label{eq:mf2x2}
\phi_1 & = & \frac{\tau_1}{b_{11}P_{11}} =
\frac{c_1}{b_{11}}\phi_1^{MF}, \ \ \ 
c_1  \equiv  \int d\,\vec{q}\frac{\tau_1^2}{P_{11}},   \nonumber \\ 
\phi_2 & = & \frac{\tau_2}{b_{22}P_{22}} =
\frac{c_2}{b_{22}}\phi_2^{MF},  \ \ \  
c_2  \equiv  \int d\,\vec{q}\frac{\tau_2^2}{P_{22}}.
\end{eqnarray} 
\noindent We remark that the MTXF is a non-symmetric matrix in general.
The variances of the two filtered maps are
\begin{eqnarray} \label{eq:variances12}
\sigma^2_{w_1} & = & \frac{1}{Nb_{11}},  \nonumber \\  
\sigma^2_{w_2} & = & \frac{1}{Nb_{22}}.
\end{eqnarray}
\noindent From the last equation, one obtains

\begin{equation}
\frac{\sigma^2_{w_1}}{\sigma^2_{w_2}} = \frac{b_{22}}{b_{11}}. 
\end{equation}

\begin{figure}
  \includegraphics[width=\columnwidth]{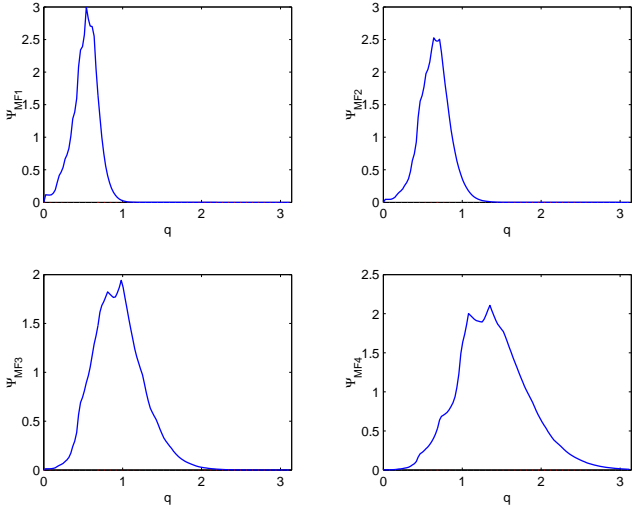}
  \caption{Standard matched filters, in Fourier space, for the four
  channels in one of the patches considered. \label{fig:matchedf}}
\end{figure}

\begin{figure*}
  \includegraphics[width=\textwidth]{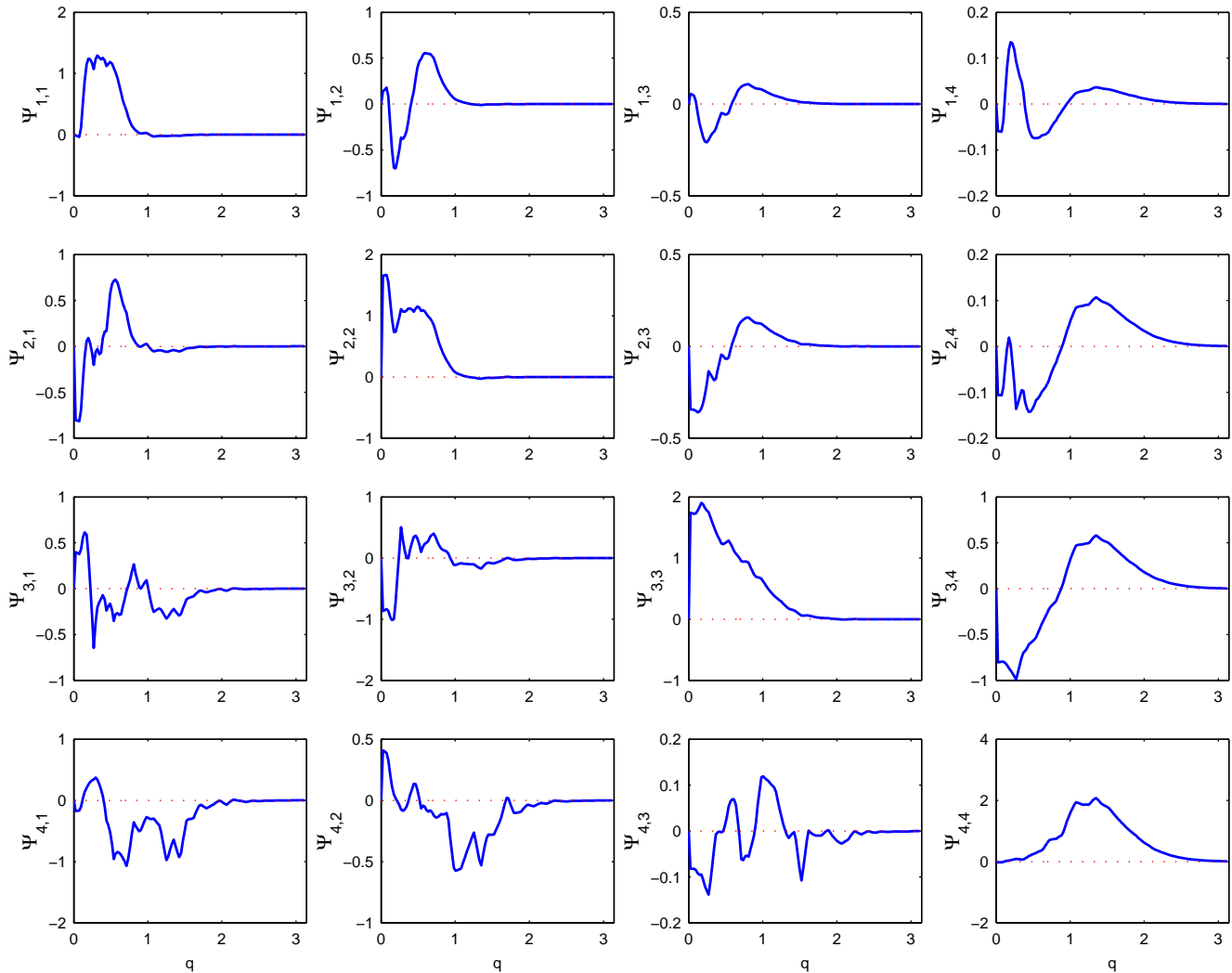}
  \caption{The different elements of the $4\times 4$ matrix of
  filters, in Fourier space, for the same patches as in
  figure~\ref{fig:matchedf}. 
  \label{fig:matrix4x4}}
\end{figure*}

\subsubsection{Relative gain of the different channels} \label{sec:relative}

An useful quantity that can be used to assess the performance of a
filter is its \emph{gain} factor, that is, the increment of the signal
to noise ratio that a fiducial source experiments thanks to the
application of a filter. Of course, gain is not a universal quantity
of the filter, since it depends on the statistical properties of the
background noise, but given a particular image the relative gain of
two filters provides a good intuitive measurement of their respective
performances. For unbiased filters (i.e. filters that preserve the
amplitude of the signal) the gain is just the ratio of noise
dispersions before and after filtering:
\begin{equation}
G=\frac{\sigma}{\sigma_w},
\end{equation}
\noindent
where $\sigma$ is the noise dispersion before filtering and $\sigma_w$
is the noise dispersion after filtering. 

When comparing the performances of matched matrix filters with those
of standard matched filters,~\citet{herranz08a} observed an apparent
trend: in all the few cases they studied, the gain factor of the
matrix filters was similar to the one of matched filters for at least
one of the channels, and significantly higher for the
others. Moreover, the channel with the lower gain was the one with the
worse signal to noise ratio after filtering. At the moment, it was not
clear if this was a universal trend of just a result obtained by
chance given the low number of cases under studied in that work. As we
will see below, the results of this work seem to support this observed
trend.

There is a qualitative argument that sheds light over this phenomenon:
for the sake of simplicity let us consider two channels and identical
profiles $\tau_1 = \tau_2$. If one assumes that $P_{11}< P_{22}$ then
the equations (\ref{eq:mf2x2},\ref{eq:variances12}) lead to
\begin{equation}
\sigma_{w_1} < \sigma_{w_2},\ \ \  
\sigma^{MF}_{w_1} < \sigma^{MF}_{w_2},
\end{equation}
\noindent for the MTXF and standard MF, respectively. Therefore, the
original map with less variance gains more with either filter. On the
other hand, assuming that $P_{12} \ll P_{11}< P_{22}$, taking into
account equations (\ref{eq:mf2x2},\ref{eq:variances12}) and the
Schwartz inequality, one obtains
\begin{equation}
\sigma_{w_i} \leq \sigma^{MF}_{w_i},
\end{equation}
\noindent i.e. the MTXF outperform the standard MF.

\section{Application to Planck radio sources} \label{sec:toplanck}

In this section, we will describe an application of the new
multiwavelength filtering technique to realistic simulations of the
sky as it will be observed by the \emph{Planck} mission. As a example,
we will focus on the blind detection of extragalactic radio
sources. Therefore we will simulate the 30, 44, 70 and 100 GHz
\emph{Planck} channels. Note that although a simultaneous
multiwavelength filtering of the nine \emph{Planck} channels is
relatively easy to do (it would imply the use of a $9 \times 9$ matrix
of filters, that can be easily handled by any desktop computer), the
addition of infrared channels would probably help little to the
detection radio sources. A intuitive reason for this is that the
sources that are bright in the infrared range are not necesarily the
same that dominate the radio range, and vice versa. Therefore, for
this work we prefer to treat the radio sources separately. Besides,
the low-dimensional $4 \times 4$ application we describe here is more
adequate to illustrate the technique, allowing us to show the details
of the method with a relatively small number of plots.

\subsection{The simulations} \label{sec:simulations}

Sky simulations are based on the Planck Sky
Model\footnote{http://www.apc.univ-paris7.fr/APC$\_$CS/Recherche/ \\
  Adamis/PSM/psky-en.php}~\citep[PSM,][in
preparation]{psm}, a flexible software package developed by Planck WG2
for making predictions, simulations and constrained realisations of
the microwave sky.

The CMB sky is based on a Gaussian realisation assuming the WMAP
best-fit $C_{\ell}$ at higher multipoles.

The Galactic emission is described by a three component model of the
interstellar medium comprising free-free, synchrotron and dust
emissions.  Free-free emission is based on the model of~\citet{dick03}
assuming an electronic temperature of 7000 K. The spatial structure of
the emission is estimated using a H$\alpha$ template corrected for
dust extinction.  Synchrotron emission is based on an extrapolation of
the 408 MHz map of~\citet{has82} from which an estimate of the
free-free emission was removed.  A limitation of this approach is that
this synchrotron model also contains any dust anomalous emission seen
by WMAP at 23 GHz.  The thermal emission from interstellar dust is
estimated using model 7 of~\citet{fink99}.

Point sources are modeled with two main categories: radio and
infra-red. Simulated radio sources are based on the NVSS or SUMSS and
GB6 or PMN catalogues. Measured fluxes at 1 and/or 4.85 GHz are
extrapolated to PLANCK frequencies assuming a distribution in flat and
steep populations.  For each of these two populations, the spectral
index is randomly drawn within a set of values compatible with the
typical average and dispersion.  Infrared sources are based on the
IRAS catalogue, and modelled as dusty galaxies. In addition, the
emission of a large number of blended infrared galaxies, not present
individually in the IRAS catalogue, is simulated to model the Far
Infrared Background~\citep{gnuevo05}.  We also include in the model a
map of thermal SZ spectral distortion from galaxy clusters, based on a
cluster catalogue randomly drawn using a mass-function compatible with
present-day observations.

\begin{figure*}
  \includegraphics[width=\textwidth]{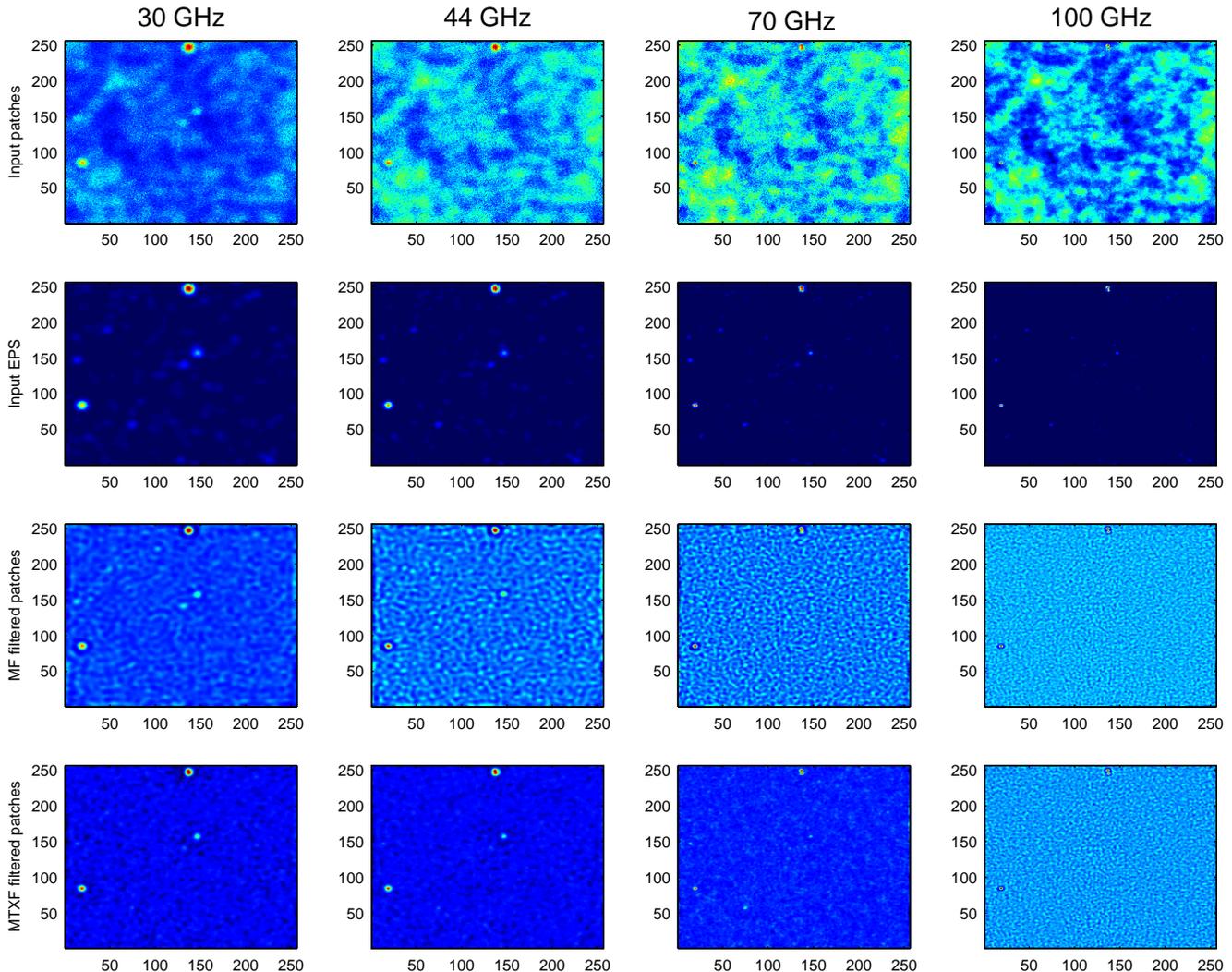}
  \caption{One of the regions of the sky. The patch is centered in the
    North Galactic Pole. The first row of images, starting from the
    top, shows the input patches at 30, 44, 70 and 100 GHz (from left
    to right of the figure). The second row of images shows separately
    the EPS contribution to the input maps, for the same
    frequencies. The third row shows the patches after having been
    filtered with the standard matched filter corresponding to each
    frequency. The bottom row shows the combined filtered maps
    resulting from the application of the matrix filters. All maps are
    expressed in MJy/sr units.
  \label{fig:patches}}
\end{figure*}

\subsection{The code} \label{sec:code}

In this work we have a used a code that reads in four all-sky FITS
maps with HEALPix~\citep{healpix} resolution parameter NSIDE=1024, one
per frequency between 30 and 100 GHz. Second, according to some
parameters given in an input file, the code uses the CPACK
libraries\footnote{http://astro.ic.ac.uk/$\sim$mortlock/cpack/} to divide
the sky in a sufficient number of overlapping flat patches such that
the $100\%$ of the sky is covered. In this pixelization scheme, we
have produced 371 patches per frequency, 14.656 square degrees and
$256\times 256$ pixels each. Then the code proceeds to run either the
MF or the MTXF algorithms on every set of four patches corresponding
to the same region in the sky. Afterwards, once the optimally filtered
image has been produced, the code looks for maxima in it producing a
subcatalogue of detections. Finally, a combined catalogue is produced,
removing possible repetitions inside a 1 FWHM radius.

\section{Results} \label{sec:results}

As described above, we have divided the sky into square flat patches
by projecting the HEALPix~\citep{healpix} maps into the tangent plane
at a set of coordinates that are regularly distributed on the
sphere. For each resulting patch we have four square images, one per
frequency channel (30, 44, 70 and 100 GHz). Then we have filtered
simultaneously the four images with the filters
(\ref{eq:matrix_filters_eqs}) specifically calculated for that region
of the sky. In paralel, we have filtered each one of the images
separately with its corresponding matched filter. Therefore, for each
input image we have two output filtered images, one obtained with the
standard matched filter and other obtained by the matrix filters as in
equation (\ref{eq:filtered_field}). Then we have applied the same
thresholding detection criterion to the two cases: for any given
filtered map, we have selected all the peaks that have at least three
connected pixels with flux above a given number of times the $\sigma$
level\footnote{Where $\sigma$ is the standard deviation of the
filtered map, excluding the region of the borders of the image; note
that this $\sigma$ level corresponds to a different flux threshold for
different filters and for different regions of the sky} of the
filtered map. Unless otherwise noted, all the plots that will be shown
in this section will refer to detections above the $5\sigma$ detection
threshold.  The detections thus obtained for the different patches
have been combined into a single whole-sky catalogs for each of the
four frequency channels and the two filtering schemes.

Figure~\ref{fig:mask} shows the Galactic mask we apply for the
analysis of the results. The mask is similar to the WMAP Kp2 mask and
it covers a highly contamined region around the Galactic plane plus a
set of irregular areas that mask other highly contamied areas of the
sky such as the Magellanic Clouds, the Ophiuchus Complex and the
Orion/Eridanus Bubble. In total, we are masking $14.91\%$ of the
pixels of the sky.

\subsection{Detail of a single patch}

Before discussing the results for all the sky outside the mask just
described above, let us illustrate the qualitative functioning of the
filters taking as a example just one sky patch. In the first row of
plots of figure~\ref{fig:patches} we show the aspect of sky in the
first of the patches we have studied, centered in the Galactic North
Pole. It is a region of the sky with a very low contamination from
Galactic emission, with two point sources that are clearly visible to
the naked eye (at least at 30 GHz), but many others are hidden amid
the diffuse components. EPS alone are shown in the second row of plots
of the figure.

Figure~\ref{fig:matchedf} shows the matched filters, in Fourier space,
for the four different channels shown in
figure~\ref{fig:patches}. Figure~\ref{fig:matrix4x4} shows the 16
elements of the corresponding matrix of filters, also in Fourier
space. Note that the filters in the diagonal look roughly similar to
the matched filters whereas the off-diagonal elements are quite
different. This can be intuitively explained in the following way: the
diagonal element $\Psi_{kk}$ is designed to produced a maximum
contribution of the source profile $\tau_k$ in the map $D_k$ whereas
the $\Psi_{kl}$, $l\neq k$ element is designed to produce a minimum
contribution of the source profile $\tau_l$ in the map $D_k$. This way
the off-diagonal elements of the filtering contribute to reduce noise
but do not introduce bias in the determination of the fluxes in the
$k^{\mathrm{th}}$ map.

The third and fourth rows of figure~\ref{fig:patches} show the output
filtered patches for the matched filter and the matrix filters,
respectively. Note that for the 44 and 70 GHz channels the output
matrix filtered maps look far cleaner than their matched filtered
equivalents. For the 30 GHz channel the distinction is not so clear
(the matrix filtered image looks cleaner, but some of the sources that
are easily visible in the matched filtered image are apparently
missing; we will see later that this is only a visual
effect). Finally, for the 100 GHz channel both filtered images look
practically identical. 

The gain factors obtained for these images with the MTXF are
$[2.9,3.8,3.5,2.8]$ for the $[30,44,70,100]$ GHz channels. The gain
ratio between the MTXF and the MF are $G_{MTXF}/G_{MF} =
[1.38,1.52,1.49,1.00]$ for the $[30,44,70,100]$ GHz channels.  Thus,
the MTXF-filtered image with lower gain (100 GHz) is the one that is
more similar to the correspondent MF-filtered image. Besides, the 100
GHz map is the one with higher variance before filtering.  We will see
in the next section that the all-sky results confirm this rule.

\subsection{All-sky results}

We can use the knowledge on the input EPS we have simulated to control
the quality of our catalogs in terms of number of true and false
detections. The criterion we use to decide whether a detection is a
true one or a false one is purely positional: an object of the catalog
is considered a true detection if it is closer than a certain matching
radius $r$ of a input source and a false (spurious) detection
otherwise. For this work we use a matching radius $r=2R_0$, where
$R_0$ is the width of the Gaussian beam correspondig to the channel
under study. This radius is smaller than one FWHM, so the criterion we
use is quite stringent. Considerations about the flux matching will be
made later in section~\ref{sec:fluxes}.

\subsubsection{Number of detections} \label{sec:numdets}

\begin{figure}
  \includegraphics[width=\columnwidth]{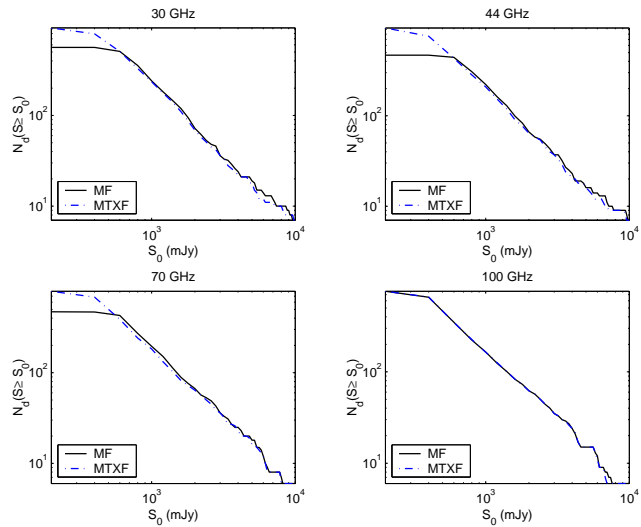}
  \caption{Number of true detections with true fluxes above a given
  flux value. Black solid line: detections with the matched
  filter. Blue dot-dashed line: detections with the matrix filters.
  \label{fig:deflux}}
\end{figure}

Figure~\ref{fig:deflux} shows the number of detected sources in our
$5\sigma$ catalogs that have true fluxes $S>S_0$, as a function of the
flux threshold $S_0$. The number of detections obtained with the MF is
shown by a solid black line, whereas the number of detections obtained
with the MTXF is shown by a dot-dashed blue line (colors are available
only in the online version of the paper).

Two different cases can be observed: the 100 GHz channel and the other
three channels. At the 100 GHz channel the performance of the two
kinds of filters is almost identical.  From 30 to 70 GHz and high and
intermediate fluxes ($\geq 0.6$ Jy) the two methods detect very
similarly. The MTXF curve runs slightly below the MF curve: the
difference consists of a few high flux sources that are somehow missed
by the MTXF. We will discuss this problem in more detail below.

In the low flux region of the plots the number of detected sources
stop growing and the curves reache a plateau. The knee point of the
curves roughly indicates the detection limit of the $5\sigma$
catalogs. At 30, 44 and 70 GHz this turnover point occurs around 600
(400) mJy for the matched (matrix) filters.  In other words, the MF
reaches its detection limit at fluxes higher than the MTXF. Therefore
the MTXF are able to go deeper and to detect many more faint sources
than the MF. At 100 GHz both filters have their turnout points located
around 400 mJy.

We have checked that the number of $5\sigma$ detections obtained with
the MF roughly agrees (taking into account the different sky coverage)
with the results obtained in previous works~\citep{can06,challenge08}
that have made use of similar Planck simulations.

\subsubsection{Completeness} \label{sec:completeness}

Figure~\ref{fig:completeness} shows the completeness level as a
function of the flux of the $5\sigma$ catalogs obtained with the MF
and the MTXF. Here the completeness is defined, as usual, as the ratio
between the number of recovered true sources and the total number of
input sources over a given flux limit. The $95\%$ completeness level
is marked by a horizontal green dotted line. The $95\%$ completeness
fluxes are, for the matched (matrix) filters, the following: at 30
GHz, 610 (540) mJy; at 44 GHz, 460 (340) mJy; at 70 GHz, 390 (270) mJy
and, at 100 GHz, 270 (270) mJy. Three issues about
figure~\ref{fig:completeness} deserve detailed comments:

\begin{figure}
  \includegraphics[width=\columnwidth]{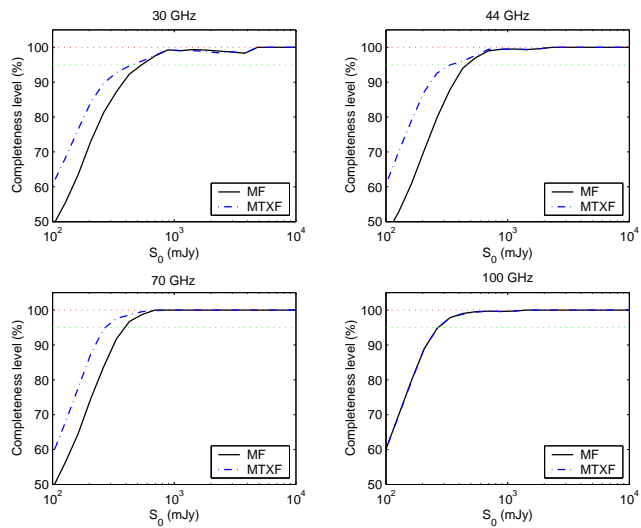}
  \caption{Completeness diagram. Black solid line: matched
  filter. Blue dashed line: matrix filters. The $100\%$ and the $95\%$
  completeness levels are marked by a red dotted and a green dotted
  horizontal lines, respectively. 
  \label{fig:completeness}}
\end{figure}

If compared to the data in Table 2 of~\citet{can06}, our present
completeness limits are higher than theirs. This is due to two causes:
on the one hand, in this work we include regions of the sky that are
closer to the Galactic plane and therefore more contamined (and
therefore, the $5\sigma$ detection threshold corresponds to a higher
flux), where it is more likely to loss some sources, and, on the other
hand, our present detection/selection criterion requires to have peaks
with at least three connected pixels. This more stringent criterion
helps to reduce the number of spurious detections, but ocasionally
makes us loss some true sources as well.

Regarding the different channels, for the case of 30, 44 and 70 GHz
the MTXF show a better completeness level at intermediate and low
fluxes. This is because MTXF do amplify better the point source signal
with respect to the foregrounds and therefore they can reach lower
detection limits. At 100 GHz, both kinds of filtering lead to the same
completeness levels.

Both methods do miss some bright sources, even at fluxes $>1$ Jy. For
example, at 30 GHz both the MF and the MTXF fail to detect 4 sources
with fluxes $>1$ Jy. Three of them are common for the two filters and
in all the three cases are sources that are in heavily contamined
regions at low Galactic latitude, very close to the border of the
mask. Apart from these three missing sources, the MF misses a source
that is detected by the MTXF and the MTXF miss a source that is
detected by the MF. The first one corresponds to a source close to the
LMC and a few pixels away from our mask. The second case corresponds
to a source that is 2.69 Jy source that is only 35 arcmin away from a
detected 3.23 Jy source. Although the distance between both sources is
slightly larger than the matching radius we used for this channel, the
MTXF is unable in this case to resolve the two sources individually.

This suggest that the mask we have used is good enough to avoid most
spurious detections, but it may be insufficient to guarantee
completeness. Besides, a problem of blending with the MTXF arises when
two high flux sources lie very close one to another. Although this
situation is not frequent, this may explain the slight loss of
performance of the MTXF in figure~\ref{fig:deflux} for high and
intermediate fluxes.

\subsubsection{Reliability} \label{sec:reliability}

Another interesting indicator of the performance of the filters is the
reliability of the catalogs obtained with them. Let it be $N_d(\nu)$
the number of true detections above a certain detection threshold
$\nu$ and $N_s(\nu)$ the number of spurious detections (false alarms)
above the same threshold. Then we define the reliability in the usual
way
\begin{equation} \label{eq:reliability}
  r(\nu) = \frac{N_d(\nu)}{N_d(\nu)+N_s(\nu)}.
\end{equation}

\begin{figure}
  \includegraphics[width=\columnwidth]{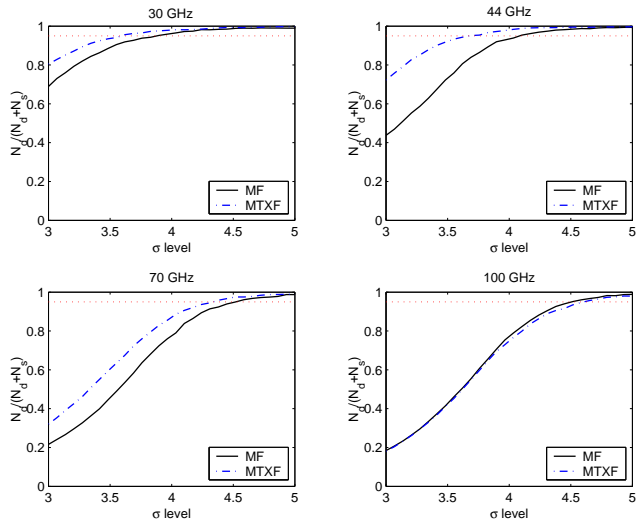}
  \caption{Reliability of the catalogs as a function of the limiting
  $\sigma$ threshold for the two filters and the four channels. Black
  solid line: MF. Blue dot-dashed line: MTXF. The red dotted line
  shows the $95\%$ reliability level.
  \label{fig:reliability}}
\end{figure}

The reliability of our $5\sigma$ catalogs is well above the $95\%$
level for the two filtering schemes and the four channels we
considered. In order to show the performance of the filter for lower
reliability levels, in figure~\ref{fig:reliability} we have gone
deeper in the detection, down to the $3\sigma$ level. The red dotted
line in the figure~\ref{fig:reliability} shows the $r=95\%$
reliability level.  For the 30, 44 and 70 GHz channels the MTXF allow
us to go to lower detection thresholds than the MF for a fixed
required reliability level. For the 100 GHz channel the situation is
the opposite, but the difference is small. Table~\ref{tb:95percent}
shows the threshold limits at the $95\%$ reliability level and the
corresponding number of true detections for the two filters and four
channels considered. The improvement of the MTXF with respect to the
MF is significant for the 30, 44 and 70 GHz channels.

\begin{table*}
 \centering
  \caption{Threshold $\sigma_{95\%}$ limit required for a $95\%$
  reliability, and corresponding number $N_{d,95\%}$ of detections for
  such a threshold.}
  \begin{tabular}{@{}ccccc@{}}
  \hline Frequency (GHz) & $\sigma_{95\%,MF}$ & $N_{d,95\%,MF}$ &
  $\sigma_{95\%,MTXF}$ & $N_{d,95\%,MTXF}$ \\ 
  \hline 
  30  & 3.9 & 900 & 3.6 & 1600 \\ 
  44  & 4.1 & 705 & 3.7 & 1550 \\ 
  70  & 4.5 & 580 & 4.3 & 1000 \\ 
  100 & 4.5 & 940 & 4.6 & 895  \\ 
  \hline
\end{tabular}
\label{tb:95percent}
\end{table*}

\subsubsection{Receiver operating characteristics} \label{sec:roc}

Yet another way to comparatively study the performance of two
detectors is the so-called \emph{receiver operating characteristic}
(ROC), or simply \emph{ROC curve}. ROC curves are profusely used in
detection theory because they provide a direct and natural way to
relate the costs/benefits of the decision making associated to the
detection process. Let us consider the two following quantities:
the \emph{true positives ratio} (TPR) is defined as
\begin{equation} \label{eq:TPR}
  \mathrm{TPR} = \frac{N_d}{NT},
\end{equation}
\noindent 
where, as before, $N_d$ is the number of true detections (true
positives) obtained for a certain detection threshold and $NT$ is the
total number of objects (in this case, simulated point sources) in the
data set. The TPR is related, but not equivalent, to the completeness
defined above (the number in the denominator of the completeness
depends as well on the detection threshold, but $NT$ does not).

The other quantity of interest is the spurious detection, false
alarm or \emph{false positive ratio} (FPR):
\begin{equation} \label{eq:FPR}
  \mathrm{FPR} = \frac{N_s}{FT},
\end{equation}
\noindent 
where $N_s$ is defined as in section~\ref{sec:reliability} and $FT$ is
the total number of candidates (in our case, peaks in the filtered
images) that can be identified or not as `detections' (true or false)
by the detector.  Therefore, the is an indirect relation between FPR
and the reliability, but note that the quantity in the denominator in
eq. (\ref{eq:FPR}) does not depend on the threshold. Both quantities,
TPR and FPR, take values in the interval $[0,1]$ and they are called
the \emph{operating characteristics} of the detector. The TPR can be
directly associated to the power of the detector and the FPR is
related to its significance.

ROC curves are constructed by plotting the fraction of true positives
(TPR) vs. the fraction of false alarms (FPR). They convey at a single
glance the same information that can be found by taking together the
figure~\ref{fig:reliability} plus a set of tables akin to
table~\ref{tb:95percent} obtained at different reliability
levels. For any fixed false alarm ratio the ROC curve tells the
(normalized) number of true detections we have. ROC curves facilitate
the comparison between two or more detectors (in our case, filters):
the curve that lies above in the plot is closer to the optimal
performance than the curves below.

\begin{figure}
  \includegraphics[width=\columnwidth]{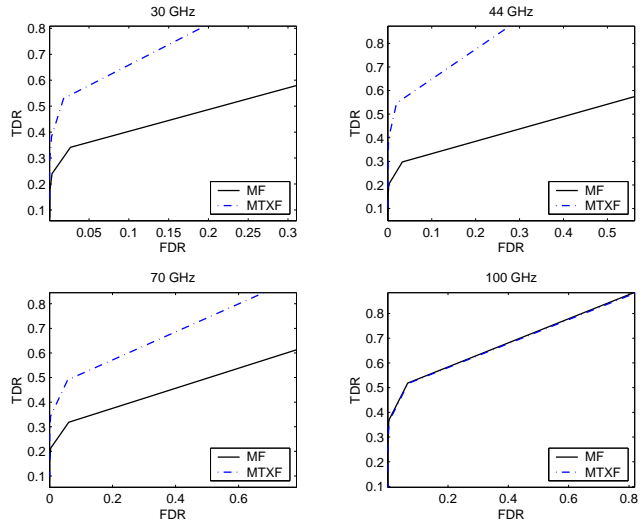}
  \caption{ROC curves for the
  filtering schemes and the four channels considered. Black solid
  line: matched filter. Blue dashed line: matrix filters. 
  \label{fig:roc}}
\end{figure}

Figure~\ref{fig:roc} shows the ROC curves for the different cases and
channels under consideration. The black solid line corresponds to the
matched filter and the blue dashed line corresponds to the matrix
filters (colors are available in the online version of the paper and
on request to the authors). The number $NT$ used for the plot is the
total number of sources that are outside the masked area of the sky
and that have fluxes above 150 mJy. The number $FT$ is the total
number of maxima found in the filtered images above the mimimum
considered threshold. The plots are made for $\sigma$ detection
thresholds in the interval $\sigma \in [3,10]$.  For this range of
detection thresholds, the ROC curve corresponding to the matrix
filters is clearly above the one corresponding to the matched filters
for the 30, 44 and 70 GHz channels, meaning that if we fix any
required spurious detection ratio we always have more true detections
with the matrix filters than with the matched filters.  For the 100
GHz channels the two curves run practically in parallel, but the
matched filters are slightly above the matrix filters. This is related
to the missing sources problem described in
section~\ref{sec:completeness}.

\subsubsection{Flux estimation} \label{sec:fluxes}

\begin{figure}
  \includegraphics[width=\columnwidth]{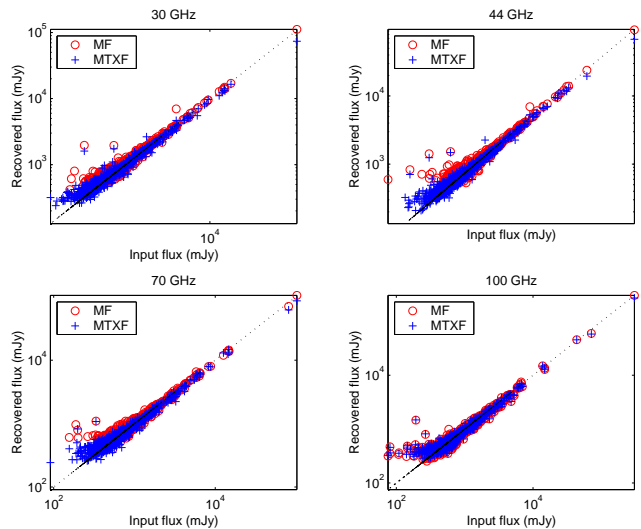}
  \caption{Estimation of the flux for the two filtering
  techniques. Red circles: matched filter estimation. Blue crosses:
  matrix filter estimation.
  \label{fig:fluxes}}
\end{figure}

Figure~\ref{fig:fluxes} shows how both kinds of filters do recover the
fluxes of the sources in all the considered cases. Circles represent
fluxes recovered with the MF and crosses represent fluxes recovered
with the MTXF. There is an excellent agreement between true input and
recovered fluxes for all the cases and between fluxes obtained with
the MF and the MTXF. At low fluxes, the MF estimates show the
well-know selection Eddington bias before than the MTXF. This is
particularly evident at 44 and 70 GHz, but also visible at 30 GHz. As
seen before, the MTXF filtered maps have less noise than the maps
filtered with the MF and therefore the selection bias appears at a
lower level. This also manifests in the smaller dispersion of blue
crosses around the true value with respect to the red
circles. However, for very high fluxes it seems that MTXF tends to
underestimate the flux of the sources. At 100 GHz both filters lead to
virtually the same results.

\begin{figure}
  \includegraphics[width=\columnwidth]{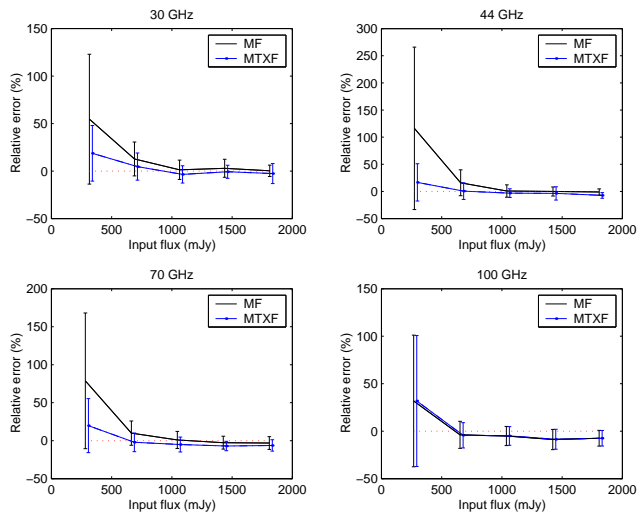}
  \caption{Relative flux estimation error and its dispersion. Black
  lines: matched filters. Blue solid line plus dots: matrix filters.
  For clarity, the lines corresponding to the matrix filter have been
  slightly displaced towards the right.
  \label{fig:rel_flux_error}}
\end{figure}

Figure~\ref{fig:rel_flux_error} shows the relative flux error for the
two filters and the four considered channels. We define the relative
flux error as
\begin{equation} \label{eq:rel_err}
  e_{rel} = 100 \times \frac{\hat{S}-S_0}{S_0},
\end{equation}
\noindent
where $\hat{S}$ is the flux we estimate with the filters and $S_0$ is
the input flux. We take five flux bins between the flux of the
faintest detected source and 2 Jy and for each bin we compute the
average value and the dispersion of the relative error. The lowest bin
is dominated by Eddington bias. This bias can be quite large, as is is
the case for the MF at 44 GHz ($115\%$ bias). The MTXF suffer less
Eddinton bias than the MF (except, as usual, for the 100 GHz case,
where the two filters behave very similarly). As mentioned before, the
dispersion around the mean value is smaller for the MTXF. Finally, it
is worth noticing the small negative bias suffered by both filters at
high fluxes. This bias is not evident at 30 GHz, but it is easier to
detect at higher frequencies. MTXF seem to be more biased ($-6\%$,
$-6\%$ and $-8\%$ at 44, 70 and 100 GHz, versus the MF biases that are
$-1\%$, $-3\%$ and $-8\%$ for the same frequencies). This kind of
effect at high fluxes has been noticed before in related filtering
applications~\citep{herr02c} and it is usually attributed to the
non-ideality of the pixelized data, that makes imperfect the process
of calculation of the normalization of the filters. In any case, the
bias is relatively small and can be calibrated by means of
simulations.

\section{Conclusions} \label{sec:conclusions}

Although there is a relatively large number of works in the literature
devoted to the detection of point sources in CMB images, there is
practically no one that addresses the problem from the multi-frequency
point of view. The reason is that each individual extragalactic point
source has its own (a priori unknown) spectral behaviour and this
makes it very hard to accomodate them in classic component separation 
schemes. 

In this work we apply a novel linear filtering technique, the
\emph{`matched matrix filters'}, introduced
by~\citet{herranz08a}. Matched matrix filters incorporate full spatial
information, including the cross-correlation among channels, without
making any a priori assumption about the spectral behaviour of the
sources. The basic underlying idea is that in all the considered
channels the sources do appear in the same unknown positions (but with
different, unknown intensities) and have known spatial profiles given
by the experiment point spread function, while some components of the
background (i.e. CMB and Galactic emission) are correlated among
channels. Then a clever linear filtering/combination of images can
lead to a substantial reduction of the background, and therefore to
lower source detection thresholds.  The resulting expresion of the
filters takes form of a matrix given in terms of the cross-power
spectra and the profile of the sources in the different maps.

We describe in detail the formalism of the matched matrix filters,
looking in detail into some particular cases of interest, and we apply
them to the detection of radio sources in realistic all-sky
\emph{Planck} simulations at 30, 44, 70 and 100 GHz. We use
state-of-the-art simulations developed by the \emph{Planck} WG2
Collaboration that include all the astrophysical components and
instrumental specifications of the \emph{Planck} mission. In order to
tackle the strong non stationarity of the sky emission, we divide the
sky into 371 $14.656\times 14.656$ overlapping square degrees flat
patches where we apply the two filtering techniques. For each patch
and frequency we obtain a sub-catalog of detections. Then all the
sub-catalogs corresponding to the same frequency are combined into an
all-sky catalog. In order to compare with a well stablished
mono-frequencial approach, we repeate the same process using the
standard matched filter.

We compare both methods in terms of reliablity, completeness, receiver
operating characteristics and flux accuracy. We find that for the
three lower frequency channels (30, 44 and 70 GHz) the new matched
matrix filters clearly outperform the standard matched filters for all
these quality indicators. The matched matrix filters decrease
significantly the noise level, what translates into a lower detection
threshold and a reduced number of false detections. The flux
estimation is consequently improved, with a lower dispersion around
the true input value and a lower flux at which the well-know selection
Eddington bias occur. The improvement is particularly evident at 44
GHz. At 100 GHz, however, the performance of the two filters is very
similar. We indicate some possible reasons for this behavior, based on
general analytical considerations about the structure of the filters.
 
One of the most interesting ways to compare two catalogs obtained with
different methods is to set a fixed level of reliability, cut the
catalogs at the corresponding points and compare how many detections
there are in each of them. We find a noticeable increment of the
number of true detections for a fixed reliability level obtained with
the matched matrix filters with respect to the standard matched
filters. In particular, for a $95\%$ reliability we practically double
the number of detections at 30, 44 and 70 GHz: the ratio between the
number of detections obtainted with the MTXF and the MF for these
channels are 1.8, 2.2 and 1.7, respectively. 

We would like to stress once more the importance of including
multi-frequency information in this approach. The new matched
multi-filters can be used to increase very significatively the number
of extragalactic point source detections in upcoming CMB experiments
such as \emph{Planck} as well as in current experiments such as
WMAP. A work on the application of this technique to the 5 year WMAP
data is in preparation.

Although here we have tested the MTXF to the particular case of the
detection of radio sources in the low frequency channels of
\emph{Planck}, the same technique can be easily applied to other
frequency bands, or to other fields of image analysis where pointlike
objects appear in different frames (images).

\section*{Acknowledgments}

The authors acknowledge partial financial support from the Spanish
Ministry of Education (MEC) under project ESP2004-07067-C03-01 and
from the joint CNR-CSIC research project 2006-IT-0037. MLC
acknowledges the Spanish MEC for a postdoctoral fellowship. JLS
acknowledges partial financial support by the Spanish MEC and thanks
the CNR ISTI in Pisa for their hospitality during his sabbatical
leave. Partial financial support for this research has been provided
to JGN by the Italian ASI (contracts Planck LFI Activity of Phase E2
and I/016/07/0 `COFIS') and MUR. We acknowledge the use of the
HEALPix~\citep{healpix} code for all sphere-based computations. The
authors acknowledge the use of the Planck Sky Model, developed by the
Component Separation Working Group (WG2) of the \emph{Planck}
Collaboration.

\bibliographystyle{mn2e}
\bibliography{herranz_bib}

\label{lastpage}

\end{document}